\documentclass[aps,twocolumn]{revtex4}
\usepackage{graphicx}

\def\13g{$^{13}\Sigma_g^+$}
\def\bea{\begin{eqnarray}}
\def\eea{\end{eqnarray}}
\def\be{\begin{equation}}
\def\ee{\end{equation}}

\begin{document}

\title{Collisional properties of trapped cold chromium atoms}
\author{Zoran Pavlovi\'{c}$^{1}$}
\author{Bj\"{o}rn O. Roos$^{2}$}
\author{Robin C\^{o}t\'{e}$^{1}$}
\author{H. R. Sadeghpour$^{3}$}
\affiliation{$^{1}$Physics Department, University of Connecticut,
2152 Hillside Rd., Storrs, CT 06269-3046}
\affiliation{$^{2}$Department of Theoretical Chemistry, 
Chemical Center, P.O.B. 124 S-221 00 Lund, Sweden}
\affiliation{$^{3}$ITAMP, Harvard-Smithsonian Center for Astrophysics, 60
Garden Street, Cambridge, MA 02138.}

\date{\today}
\begin{abstract}
We report on calculations of the elastic cross section and thermalization rate
for collision between two maximally spin-polarized chromium 
atoms in the cold and ultracold
regimes, relevant to buffer-gas and magneto-optical cooling of chromium atoms.
We calculate {\it ab initio} potential energy curves for Cr$_{2}$
and the van der Waals coefficient $C_{6}$, and construct 
interaction potentials between two colliding Cr atoms. 
We explore the effect of shape resonances on elastic 
cross section, and find that they dramatically 
affect the thermalization rate. Our calculated value for the $s$-wave scattering length is compared in magnitude with a recent 
measurement at ultracold temperatures.
\end{abstract}

\maketitle
 
Collisions of atoms at ultracold temperatures 
have received considerable attention 
because of their importance in cooling and 
trapping of atoms \cite{nobel-97}
and their role in high precision spectroscopy \cite{rev-julienne}
and Bose-Einstein condensation \cite{bec}.
Processes occurring in the cold regime are 
sensitive to the details of the interaction 
potentials between the colliding systems over 
an extended range of internuclear separations.
Recent experiments with chromium \cite{jabez,pfau,pfau2,doyle} 
emphasize the need for theoretical studies of Cr scattering properties. 
The interest in cooling Cr stems from its particular properties;
in its ground state $^7S_3$, it possesses a very large 
magnetic moment, 6 $\mu_{B}$ ($\mu_{B}$: Bohr magneton), making 
it an ideal atom for buffer-cooling in a purely magnetic 
trap \cite{doyle}, as well as for magneto-optical 
trapping \cite{jabez}.
In addition, anisotropic long-range interactions, such as
chromium's magnetic dipole-dipole interactions, may lead 
to  novel phenomena in BECs \cite{shlyapnikov,baranov}.  
The existence of a stable fermionic isotope, $^{53}$Cr, opens 
the possibility of obtaining fermionic degenerate
gas using sympathetic cooling. 
Chromium was also used in a new cooling scheme \cite{pfau}, 
where ultracold
chromium atoms were loaded from a MOT into 
an Ioffe-Pritchard 
magnetic trap, and cooled below $100\mu K$. However, a not so-desirable byproduct of 
large-spin collision is inelastic ``bad" scattering rates that 
deplete the trap. 

On the theoretical front, the electronic spectrum of the Cr$_2$ dimer 
poses a considerable 
numerical challenge. Chromium is the first atom in the 
periodic table with a half-filled $d$ shell
(the ground electronic configuration is Cr($3d^54s, ^7S$)) 
and the Cr$_2$ dimer is one of the most extreme 
cases of multiple metal-metal bonding. To date, 
the best attempt to calculate its interaction
potential curves is a 
multiconfiguration second-order perturbation 
theory with complete active space self-consistent 
field (CASSCF/CASPT2) \cite{cas,roos}. While some 
information on the spectroscopy of the ground 
electronic state ($^1\Sigma_g^+$ molecular symmetry) 
exists, there is no data available for the
interaction of two maximally spin-stretched Cr 
atoms in the $^{13}\Sigma_g^+$ molecular symmetry, 
i.e. total spin, $S=6$. 

In this communication, we explore the collisional
properties of Cr atoms in the cold and ultracold 
temperature regimes by revisiting the electronic 
structure of the dimer. More accurate Born-Oppenheimer 
potential energy curves dissociating to 
two ground Cr atoms were obtained. 
The van der Waals interaction coefficient, $C_6$, was 
obtained semi-empirically 
from available bound atomic transition matrix 
elements and photoionization cross sections 
to be $C_6 = 745 \pm 55$ a.u., where 
1 a.u. = $9.57\times 10^{-80}$ Jm$^6$.
The elastic cross section and collision rate coefficient were calculated using
the newly constructed potential curves and 
compared against two recent measurements of elastic 
rate coefficients \cite{doyle,pfau}. We also investigated 
the bound and resonance structure in the 
\13g potential energy 
curve and determined the resonance positions 
and widths as a function of rotational 
angular momenta.

\begin{figure}
 \centerline{\includegraphics[clip,width=3.25in]{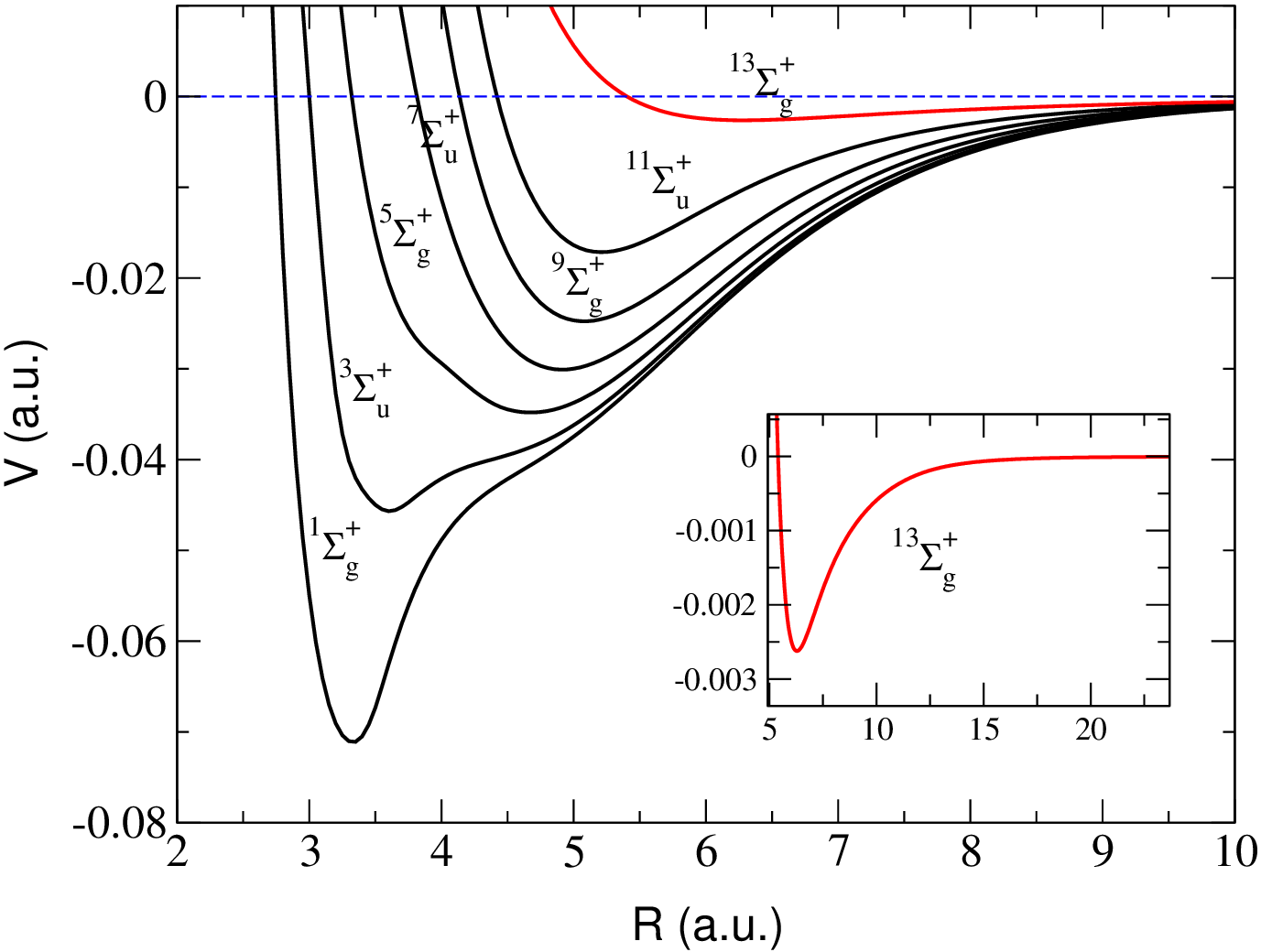}}
\caption{
    Potential curves for the ground state manifold of Cr$_{2}$ computed with the
    CASSCF/CASPT2 method. The maximally spin-streched electronic state, \13g, is shown in red and in detail in the inset.
This state supports 30 vibrational levels.
        }
\label{fig:allpot}
\end{figure}

The potential curves for Cr$_{2}$, shown in Fig.~\ref{fig:allpot},
were constructed from three
regions joined smoothly together. We first computed {\it ab initio}
potential curves using the CASSCF/CASPT2 method \cite{cas,roos}.
The CASSCF wave function is formed by distributing 12 electrons 
in the 3d and 4s derived active orbitals while keeping the 
inactive 1s, 2s, 2p, 3s, and 3p derived orbitals
occupied. The remaining dynamical electron correlation 
energy is obtained through second
order perturbation theory (CASPT2). The basis set used 
in the calculations is of the atomic
natural orbital (ANO-RCC) type contracted to $9s8p7d5f3g$. 
This basis set is relativistic and includes functions for 
correlating the $3s$ and $3p$ electrons \cite{basis}. The
Douglas-Kroll Hamiltonian was used with Fock-type correction  
$0.5*\mathbf{g}_{1}$, see \cite{gcorr}.
The full counterpoise method was used to correct energies
for the basis set superposition error (BSSE). Convergence to $10^{-10}$, in hartrees, 
was achieved, and numerical accuracy in computed 
binding energies was about $10^{-8}$. 


For separations $R\leq R^{\lambda}_{1}$, where $R^{\lambda}_{1}$ 
is the smallest separation of the  
{\it ab initio} data for the potential energy $V_{\lambda}(R)$, 
each curve was joined smoothly 
to the exponential form $V_{\lambda}(R)=c_{\lambda}\exp(-b_{\lambda}R)$, 
with the coefficients 
$c_{\lambda}$ and $b_{\lambda}$ determined 
by matching both the potential curve and its first 
derivative continuously at $R^{\lambda}_{1}$. 
At large values of $R$, 
the {\it ab initio} data were matched to the asymptotic form
\[
 V_{\lambda}(R) = -\frac{C_{6}}{R^{6}} + A_{\lambda}R^{\nu}e^{-\beta R} \; ,
 \label{eq:longrange}
\] 
where $C_{6}$ is the van der Waals coefficient \cite{note-c6}, 
and the parameters of
the exchange energy are determined according to Smirnov and 
Chibisov \cite{smirnov}: $\nu = \frac{7}{2I}-1$ and $\beta = 2I$, where
$I$ is the ionization energy of the atom ($I= 0.248664314 $ a.u. for Cr). 
The parameters $A_{\lambda}$ were found
by fitting the {\it ab initio} curves at separations where the exchange
energy was still considerable (e.g. $R$ between 10 and 14 a.u. for
$^{13}\Sigma_{g}^{+}$). The $C_{6}$ coefficient was calculated
using 
\bea
  C_{6} & = & \frac{3}{2}\left(\frac{e^{2}\hbar^{2}}{m}\right)^{2}
    \frac{1}{\left|E_{0}\right|^{3}}
    \left\{\sum_{i,j}\frac{f_{0i}f_{0j}}{v_{i}v_{j}(v_{i} + v_{j})} 
    \right. \nonumber \\ 
    & &  \hspace{-.45in} 
    \left. + 2\sum_{i}\frac{f_{0i}G(1+v_{i})}{v_{i}}
           + \int_{0}^{\infty}\frac{(df/d\epsilon)
              G(2+\epsilon)d\epsilon}{(1+\epsilon)}\right\}  ,
\eea
where contributions of bound-bound, bound-free, and free-free 
transitions are given by the first, second, and third term,
respectively.
This expression is derived from the London's \cite{margenau} formula,
assuming $v_{i}=1-E_{i}/E_{0}$ and $\epsilon=-E/E_{0}$,
where $E_{0}$, $E_{i}$, and $E$ are the ground, the $i^{\rm th}$ excited state
and continuum energies, respectively, $f_{0i}$ are the oscillator 
strengths pertaining to transitions to the ground states, and $df_{0E}$
accounts for the continuous spectrum. The auxiliary function $G(z)$ is 
given by
\be
  G(z)=\int_{0}^{\infty}\frac{(df/d\epsilon)d\epsilon}
       {(1+\epsilon)(z+\epsilon)} \; .
\label{galpha}
\ee
Values of oscillator strengths and energy levels for discrete transitions
were taken from the NIST Atomic Spectra Database \cite{nist:discrete} and
 Ref.~\cite{verner:discrete}.
The continuous oscillator strength, $df/d\epsilon$, was found
using Verner's \cite{verner:cross} analytic fits for partial photoionization 
cross sections. If we measure $dE$ in atomic units then
\be
  \frac{df}{dE}=\frac{1}{2{\pi}^{2}\alpha a_{0}^{\ 2}}\sigma_{\rm ph} \; ,
  \label{eq:df/de}
\ee
where $\alpha$ is the fine-structure constant, $a_{0}$ is the Bohr radius,
and $\sigma_{\rm ph}$ is the photonization cross section. 
The dimensionless $df\!/d\epsilon$ is obtained by
multiplying Eq.(\ref{eq:df/de}) by $|E_{0}|$ (also in a.u.).

We assess the quality of the semi-empirical computation of the van der Waals coefficient by satisfying two sum rules, namely, the 
zero-th and the inverse second moments,
\bea 
   S(0)& =&\sum_{i}f_{0i}= N \; \nonumber \\
   S(-2)=\alpha_{0} & = &\frac{e^{2}\hbar^{2}}{m |E_{0}|^{2}}  
   \left\{\sum_{i}\frac{f_{0i}}{v_{i}^2}+G(1)\right\} ,
\eea
where $N$ is the number of electrons (24 for Cr), and $\alpha_{0}$ 
is the static polarizability. We obtained $\sum_{i}f_{0i}= 22.3$
and $\alpha_{0}=85.00$ a.u., in agreement with  a recommended 
value $82\pm 20\% $ a.u. \cite{alpha0}). The resulting dispersion coefficient is expected to have an accuracy of 
about $7\%$, $C_6 = 745 \pm 55$ a.u.

Using these potential curves, we computed the elastic 
$\sigma^{\lambda}_{\rm el.}$ 
cross section for the maximally spin-aligned molecular 
state, i. e. the \13g state. 
[Results for other molecular states 
and also inelastic and spin relaxation cross 
sections will be provided in a future publication.]
The elastic cross section for the collision of two 
$^{52}$Cr atoms, composite bosons, expanded over 
the rotational quantum number, $l$,
is \cite{mott}
\be
   \sigma^{\lambda}_{\rm el.}(E) =\frac{8\pi }{k^{2}}\sum_{l \;\rm{even}}
   (2l+1)\sin ^{2}\delta^{\lambda}_{l}  \;,
\ee
where $E=\hbar^{2}k^{2}/2\mu$ is the kinetic energy of relative motion,
$\mu$ is the reduced mass, and $\delta^{\lambda} _{l}(k)$ is the $l$-th
scattering phase shift in electronic state $\lambda$. In the low-energy limit, the elastic cross 
section behaves as
\be
   \sigma^{\lambda}_{\rm el.}(E) \; \stackrel{E\rightarrow 0}{\longrightarrow} \;%
   \frac{8\pi a_{\lambda}^{2}}%
   {k^2a_{\lambda}^2+\left(1-\frac{1}{2}r_{\lambda}a_{\lambda}k^2\right)^2} \;,
\ee
where  scattering length $a_{\lambda}$ is determined
at the zero-energy $s$-wave limit, 
$a_{\lambda}=-\lim_{k\rightarrow 0}\frac{1}{k}\tan 
\delta_{0}^{\lambda}(k)$, while the effective range $r_{\lambda}$
can be found by fitting the cross section, from an 
integral expression \cite{mott},
or using quantum defect theory\cite{gao}.

\begin{figure}
 \centerline{\includegraphics[clip,width=3.25in]{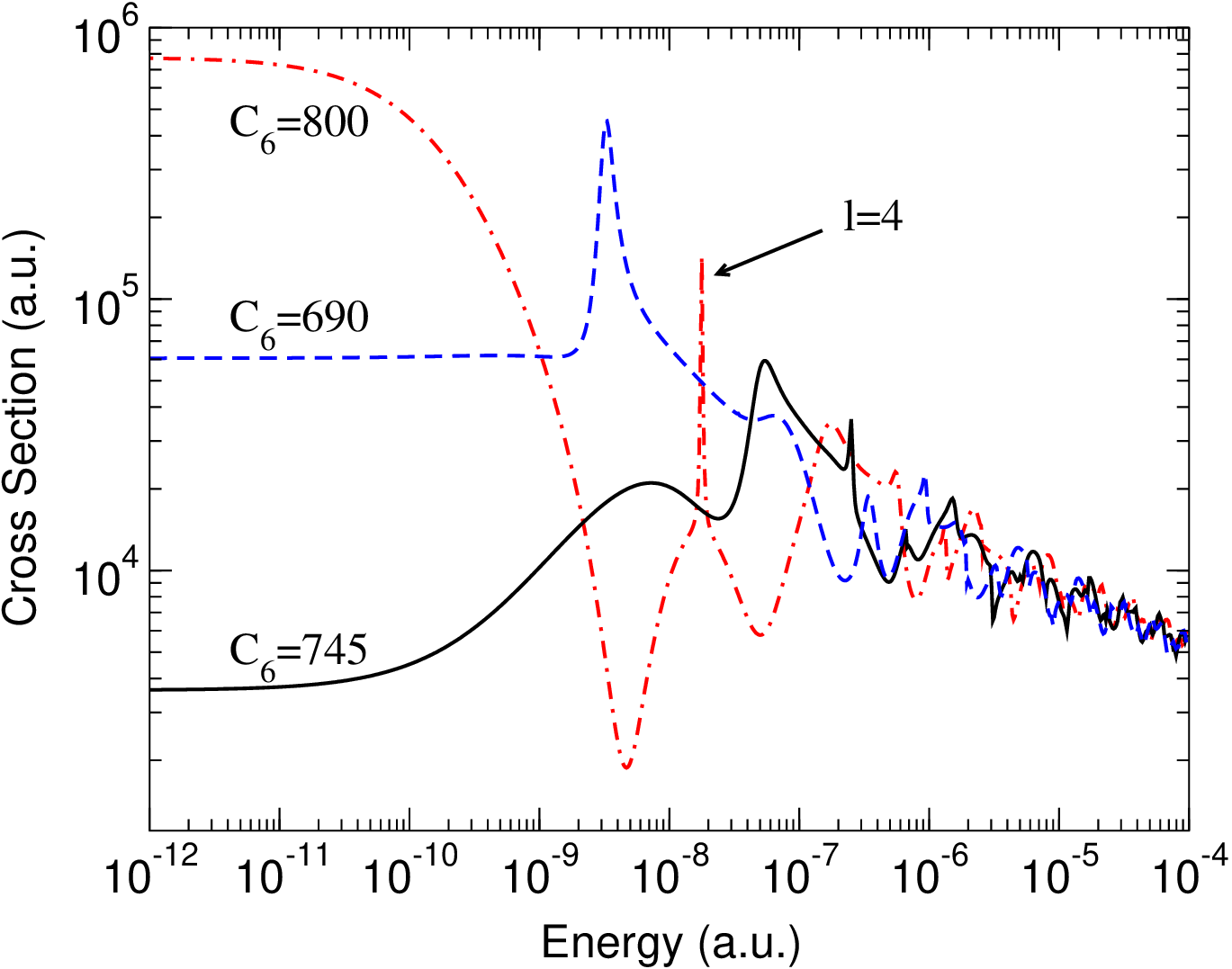}}
 \caption{
          Elastic cross sections for collision between two bosonic Cr atoms as a function of energy in the cold and ultracold 
regimes. The three curves are labeled by the value of the $C_6$ coefficient used to construct the interaction potential in each
case. See text for details. The influence of shape resonances is dramatic.
        }
\label{xsections}
\end{figure}

\begin{figure}
 \centerline{\includegraphics[clip,width=3.25in]{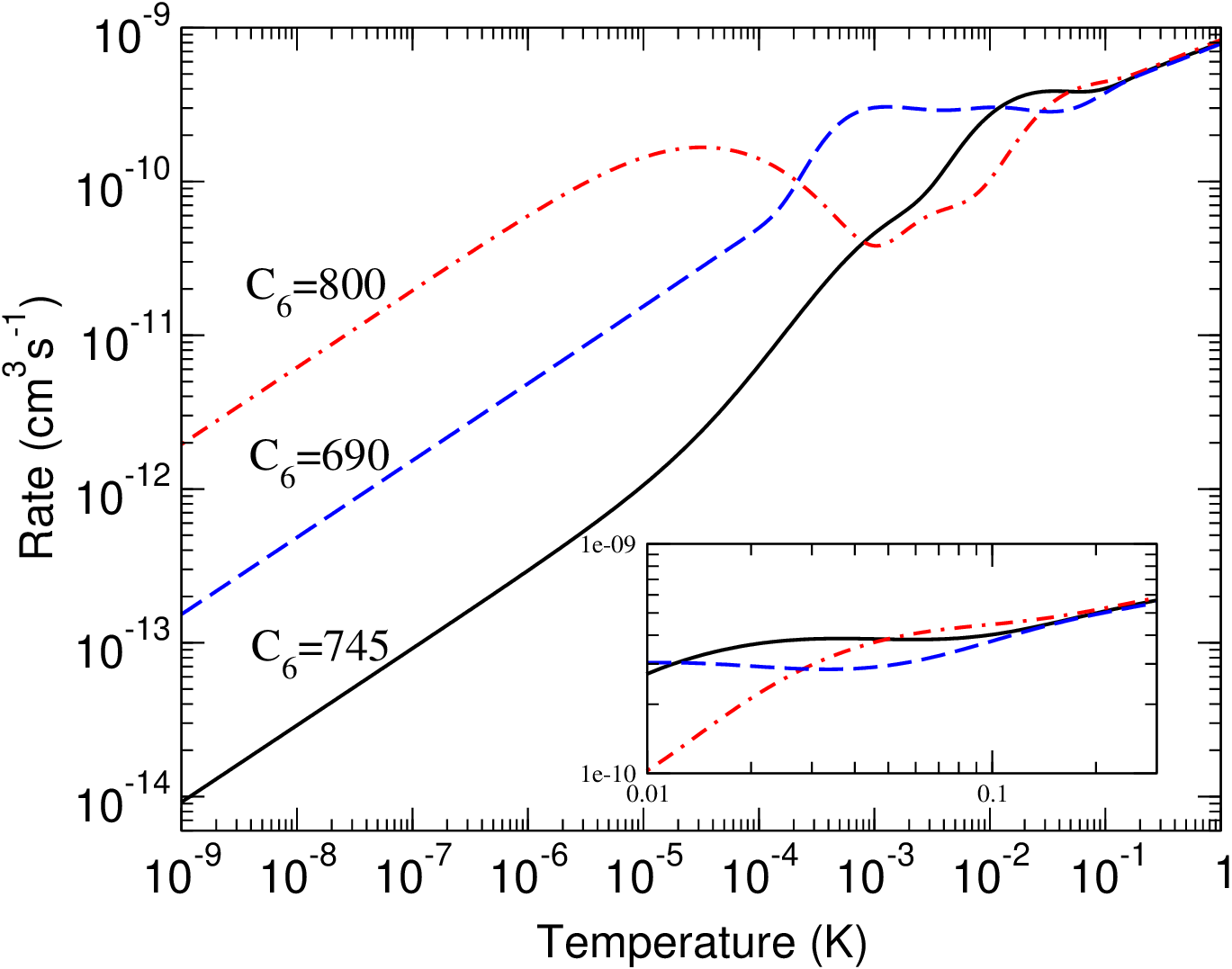}}
\caption{The calculated collision rate coefficient as a function of temperature. At temperatures above 100 mK, the 
different calculated rates are practically the same. The plateau region is shown in the inset.
}
\label{rates}
\end{figure}

The cross section as a function of the collision
energy is shown in Fig.~\ref{xsections} for three 
different \13g interaction potentials; each constructed by matching at large distances to the upper, lower, and mean
values of $C_6$. 
Note 
that at collision energies larger than $E \sim 10^{-6}$ a.u., 
aside from the shape resonance structure,  the cross sections 
differ little in magnitude, but are dramatically different in the $s$-wave limit. 
We find the scattering length for the \13g state, $a_{13}$, 
to be large and negative 
(see Table~\ref{tab1}),  
indicating that evaporative cooling should be 
efficient, but that a Bose-Einstein condensate of $^{52}$Cr  would not be stable 
for large number of atoms. This contrasts Ref. \cite{pfau} in 
which the extracted value for the 
scattering length, obtained from a fit to the experimental collision rate coefficient, Fig.~\ref{rates}, agrees in magnitude with 
our upper bound calculated value, but differs in the prediction of the sign (see also the discussion of the rates below).
Our calculated effective range expansion coefficients 
also 
agree very closely with 
the results obtained using the quantum-defect method of Gao \cite{gao}.

As the collision energy increases, the appearance of shape
resonances can lead to enormously large cross sections. 
Althougth the 
details of the resonance structure, i. e. profile and 
energy position, depend on the details of the potentials, 
the overall effect 
on the collisional rate coefficient is
only minimal at temperatures for which the shape resonances matter. 
This is portrayed in Fig. \ref{rates}
for the ultracold and cold temperatures. The rate 
coefficients for the three different potentials are similar 
for temperatures higher than 100 mK. Our rate 
coefficients are larger than those 
measured by \cite{doyle} by more 
than an order of magnitude. We do not yet 
understand the origin of this discrepancy- 
an attempt at modifying 
the potential in the long-range and in the short-range resulted 
in practically similar results. The shape and magnitude of
the calculated rate coefficient for $C_6=800$ a.u. agrees well in both 
magnitude and shape with the MOT experimental 
results,
see Fig. 3 of Ref. \cite{pfau}. 
The decline in the rate coefficient for $T > 100 \ \mu$K, as seen 
in our calculation, 
followed by a rise for $T > 1$ mK, 
is due to the appearance of the $l=4$ 
shape resonance in Fig. \ref{xsections}. The experimental 
determination of the dip
in the elastic rate could be used to infer a more 
accurate value of $C_{6}$ and hence
the scattering length as well.

\begin{figure}
 \centerline{\includegraphics[clip,width=3.25in]{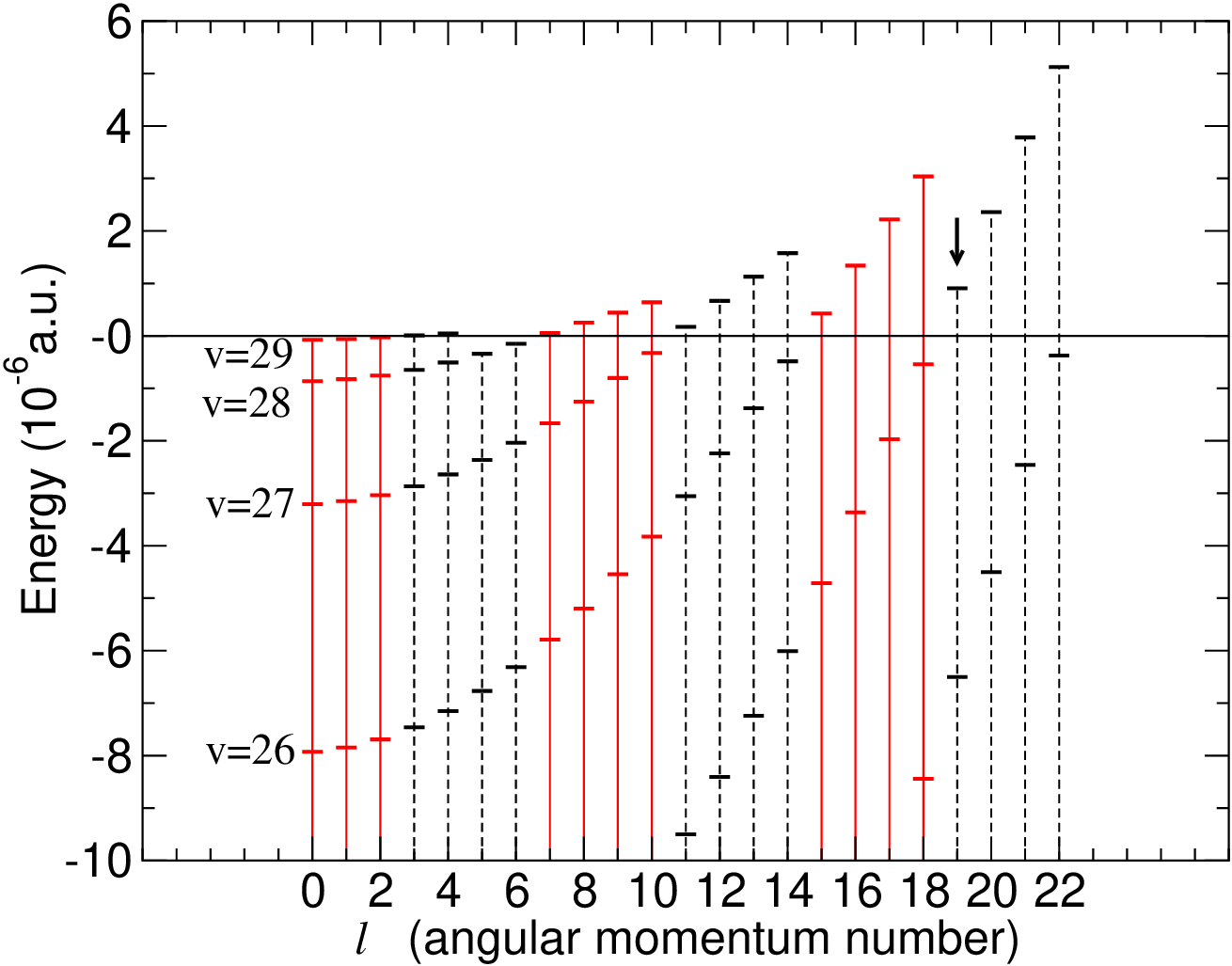}}
\caption{The bound and shape resonance structure in the 
Cr$_2$ \13g potential energy curve as a function of
the rotational quantum number $l$.
         }
\label{shape}
\end{figure}

In Fig. \ref{shape}, we give the 
bound and shape resonance structure in  
the \13g potential energy curve with $C_{6}=745$ a.u., as a
function of the rotational angular momentum, $l$. Due to 
symmetry, only even values of $l$ appear in the collision. There
are 30 bound vibrational levels for the $l=0$ partial wave. 
The grouping of the levels in solid and
dashed lines indicates the 
appearance of an additional shape resonance. 
Table \ref{tab2} gives our calculated position and  width for a number of rotational shape resonances. The narrowest 
resonance is for the $l=19$ partial wave with a lifetime of about two seconds. This resonance, however, is not populated in a 
bosonic collision. 

A plateau in the observed elastic collision rate coefficient in the range $10$ mK $< T < 300$ mK exists \cite{doyle,rob} and is
reproduced in our calculation in Fig.~\ref{rates}. It appears that this plateau is produced by the confluence of many 
collisionally-excited shape resonances in this temperature range.

We have calculated {\it ab initio} interaction potentials for collision of two Cr atoms, and obtained the long-range van der Waals 
coefficient for the first time. Elastic collision cross section and rate coefficient for two maximally-stretched chromium atoms 
have been computed in the cold and ultracold regimes and compared with available experiments. The effect of partial-wave 
shape resonances is studied. The $s$-wave scattering length and effective range have been obtained in accord with experiment. 
The disagreement with the buffer-gas cooling rate coefficient results is not fully understood and the inelastic loss rate will be 
studied in the near future. The main loss rate mechanism is expected to be spin dipole interaction.

\begin{acknowledgments}
The work of Z.P and R.C. was supported in part by the National Science Foundation grant 
PHY0140290 and University of Connecticut Research Foundation. %
The  authors would like to thank J. Doyle and R. Krems for fruitful discussions.
This work was supported by the National Science Foundation through a grant for the Institute
for Theoretical Atomic  Molecular and Optical Physics at Harvard University and Smithsonian
Astrophysical Observatory. HRS is grateful to K. Andersson for access to numerical data for an 
earlier calculation of the potential curves and to P. O. Schmidt and R. deCarvalho for valuable correspondence.
BOR thanks the Swedish 
Science Research Council (VR) for financial support.
\end{acknowledgments}

\begin{table}[ht]
\begin{center}
\begin{footnotesize}
\begin{tabular}{crr} \hline\hline \\
C$_{6}$ & a$_{13}$ & r$_{13}$\\ \hline
 & & \\
690 & 49 & 98 \\
745 & 12 & 2600 \\
800 & -176 & 213 \\
 & & \\ \hline\hline
\end{tabular}
\end{footnotesize}
\caption{ The calculated scattering length and effective range for different interaction potentials, all in atomic units.}
\label{tab1}
\end{center}
\end{table}

\begin{table}[ht]
\begin{center}
\begin{footnotesize}
\begin{tabular}{rcc} \hline\hline \\
$l$ & E (a.u) & $\Gamma$ (a.u) \\ \hline
 & & \\
 3 & 8.04(-09) & 5.82(-10)\\
 4 & 4.82(-08) & 2.87(-08) \\
 7 & 5.38(-08) & 2.03(-12)\\
 8 & 2.52(-07) & 1.60(-08) \\
 9 & 4.44(-07) & 1.04(-07)\\
10 & 6.41(-07) & 2.81(-07)\\
11 & 1.72(-07) & 1.10(-14)\\
12 & 6.66(-07) & 2.27(-09)\\
13 & 1.13(-06) & 6.11(-08)\\
14 & 1.58(-06) & 2.42(-07)\\
15 & 4.27(-07) & 6.32(-17)\\
16 & 1.34(-06) & 1.11(-10)\\
17 & 2.22(-06) & 1.82(-08)\\
18 & 3.04(-06) & 1.59(-07)\\
19 & 9.07(-07) & $\ll$ 1(-17)\\
20 & 2.36(-06) & 4.08(-12)\\
21 & 3.78(-06) & 2.70(-09)\\
22 & 5.12(-06) & 7.10(-08)\\
 & & \\ \hline\hline
\end{tabular}
\end{footnotesize}
\caption{ The rotational shape resonance parameters, position and width, 
in the \13g potential. Note that only even partial waves contribute to the 
scattering of spin polarized bosons (odd partial waves are included for completeness).}
\label{tab2}
\end{center}
\end{table}


\begin{thebibliography}{99}
\bibitem{nobel-97}
   S. Chu,
   \rmp {\bf 70}, 685 (1998);
   C.N. Cohen-Tannoudji, {\it ibid}, 707;
   W.C. Phillips, {\it ibid}, 721.
\bibitem{rev-julienne}
   J. Weiner, V.S. Bagnato, S.C. Zilio, and P.S. Julienne,
   \rmp {\bf 71}, 1 (1999).
\bibitem{bec}
  See A. J. Leggett, \rmp {\bf 73}, 307 (2001), and
  references therein.

\bibitem{jabez}C. C. Bradley, {\it et al.}, \pra {\bf 61}, 053407-1(2000).

\bibitem{pfau}P. O. Schmidt {\it et al.}, arXiv:quant-ph/0303069 v2, 2003; 
P. O. Schmidt  {\it et al.}, arXiv:quant-ph/0211032, Nov. 6 2002;
P. O. Schmidt P.O. {\it et al.}, arXiv:quant-ph/0208129v1, Aug. 20 2002.

\bibitem{pfau2} Stuhler J. {\it at al.}, \pra {\bf 64}, 031405 (2001); S. Giovanazzi, A. G\"orlitz, and T. Pfau
\prl {\bf 89}, 130401(2002).

\bibitem{doyle} J.M. Doyle {\it et al.}, \pra {\bf 65}, 021604 (2002);
  J. D. Weinstein {\it et al.}, 
\pra {\bf 57}, R3173(1998).

\bibitem{rob}Robert deCarvalho (private communication).

\bibitem{shlyapnikov}
    L. Santos, G.V. Shlyapnikov, P. Zoller, and M. Lewenstein,
 {\it Bose-Einstein Condensation in Trapped Dipolar Gases}, \prl {\bf 85}, 1791 (2000).

\bibitem{baranov} M. Baranov {\it at al.}, arXiv:condmat/0201100 (2002)

\bibitem{cas}B. O. Roos in {\it Advances in Chemical physics, ab initio methods in quantum chemistry}, Ed. K. P. Lawley,
John-Wiley \& Sons, Ltd., Chichester, 1987; B. O. Roos {\it et al.} in {\it Advances in chemical physics; new methods 
in computational quantum mechanics}, Ed. I. Prigogine and S. A. Rice, John-Wiley \& Sons, Ltd., New-York, 1996.

\bibitem{roos} B. O. Roos, and K. Andersson,
 Chem. Phys. Letters {\bf 245}, 215 ~(1995); B. O. Roos, Collect. Czech. Chem. Commun. {\bf 68}, 265(2003);
K. Andersson,
 Chem. Phys. Letters {\bf 237}, 212 ~(1995).

\bibitem{basis} The primitive basis set was: 21s15p10d6f4g. These basis sets are under 
construction for the entire periodic system (B. O. Roos, to be published).

\bibitem{gcorr}  K. Andersson,
 Theor. Chim. Acta    {\bf 91}, 31 (1995).

\bibitem{note-c6}
  Higher order dispersion coefficients, such as $C_{8}$, $C_{10}$, etc., were
  not included in our analysis.

\bibitem{smirnov}
  B. M. Smirnov, and M. I. Chibisov, Soviet Physics Jetp, {\bf 21} 624 (1965);
  E. L. Duman, and B. M. Smirnov, Optics and Spectroscopy, {\bf XXIX} 229 (1970).

\bibitem{margenau} H. Margenau,
 Phys. Rev., {\bf 56}, 1000 (1939).

\bibitem{nist:discrete}
 NIST Atomic Spectra Database : 
 \begin{verbatim}http://physics.nist.gov/cgi-bin/AtData/main_asd \end{verbatim}

\bibitem{verner:discrete} D. A. Verner,  P. D. Barthel, and D. Tytler, 
  A\&AS, {\bf 108}, 287 (1994).

\bibitem{verner:cross} D. A. Verner,  and D. G. Yakovlev,
  A\&AS, {\bf 109}, 125 (1995).

\bibitem{alpha0} CRC Handbook of Chemistry and
              Physics, 83rd Edition, Section 10-163, 
   Atomic and Molecular Polarizabilities, CRC Press, LLC, 
www.crcpress.com.

\bibitem{mott} N. F. Mott, and H. S. W. Massey,
The Theory of Atomic Collisions. Third Edition. (1965)

\bibitem{gao} B. Gao, \pra P{\bf 58}, 4222(1998).

\bibitem{dohrmann} Th. Dohrmann {\it et al.}, J. Phys. B {\bf 29}, 4461 (1996).

\bibitem{dolmatov} V. K. Dolmatov, J. Phys. B {\bf 26}, L393 (1993). 
\end{thebibliography}
\end{document}